\documentstyle[11pt,newpasp,twoside,epsf]{article}
\begin{document}

\title{Analysis of Strong Outbursts in Selected Blazars from the Mets\"ahovi
and UMRAO Monitoring Databases}
\author{Pyatunina T.B.}
\affil{Insitute of Applied Astronomy of the RAS, Zhdanovskaya St. 8, 197042
St.-Petersburg, Russia}
\author{Kudryavtseva N.A.}
\affil{St.-Petersburg State University, Bibliotechnaya sq. 2, Petrodvoretz,
198904, St.-Petersburg, Russia}
\author{Gabuzda D.C.}
\affil{Physics Department, University College Cork, Cork, Ireland}
\author{Jorstad S.G.}
\affil{Institute for Astrophysical Research, Boston University, Boston, USA}
\author{Aller M.F. and Aller H.D.}
\affil{Astronomy Department, University of Michigan, USA}
\author{Ter\"asranta H.}
\affil{Mets\"ahovi Radio Observatory, Helsinki University of Technology, Finland}

\begin{abstract}
Frequency-dependent time lags for strong outbursts in four $\gamma$-blazars
are determined. The time lags for two adjacent outbursts in 2230+114 are
correlated with the outburst amplitudes. There is evidence that bright
outbursts in 2230+114  appear with a quasi-period of ($8.0\pm0.3$)~yr.
\end{abstract}

According to both theoretical arguments (Marscher 1996; Lobanov 1998) and
observational data (Pyatunina et al. 2000; Zhou et al. 2000), two different
types of outbursts should be found in the variable radio emission of blazars,
namely, ``core'' and ``jet'' outbursts. Core outbursts show
frequency-dependent time delays, and are probably associated with
core brightening due to a primary perturbation.
Jet outbursts evolve nearly synchronously at all frequencies,
and may be associated with variability accompanying the propagation of the
perturbation along the jet. The interval between two successive core
outbursts could define the duration of an activity cycle,
from the origin of a
primary perturbation in the ``central engine" until it fades into
the quiescent jet. The questions of how stable this interval is
for a particular source and how it varies from source to source
may be key for our understanding of the activity's grand design.
In addition, frequency-dependent time delays
can be used to test models of the nonthermal
emission in blazars (Lobanov 1998; Marscher 2001).

The combined data of the University of Michigan Radio Astronomical Observatory
(UMRAO; Aller et al. 1985) and Mets\"ahovi Radio Observatory (Te\-r\"asranta
et al.  1992) provide us with radio light curves
from 4.8 to 37 GHz covering time intervals up to $\sim25$ years.
As a first study sample, we chose the four $\gamma$-ray blazars
(Jorstad et al. 2001) 0458-020, 0528+134, 1730-130 and 2230+114.

We separated the most prominent outbursts in the radio light curves
into individual components by Gaussian model fitting. The frequency-dependent
time lags $\Delta T_{max}(\nu)=T_{max}(\nu)-T_{max}(\nu=37GHz)$
for the components were determined and approximated by exponential
functions of frequency of the form (Lobanov 1998):
$\Delta T_{max}(\nu)\propto\nu^{-\alpha}$.
\begin{figure}
\plottwo{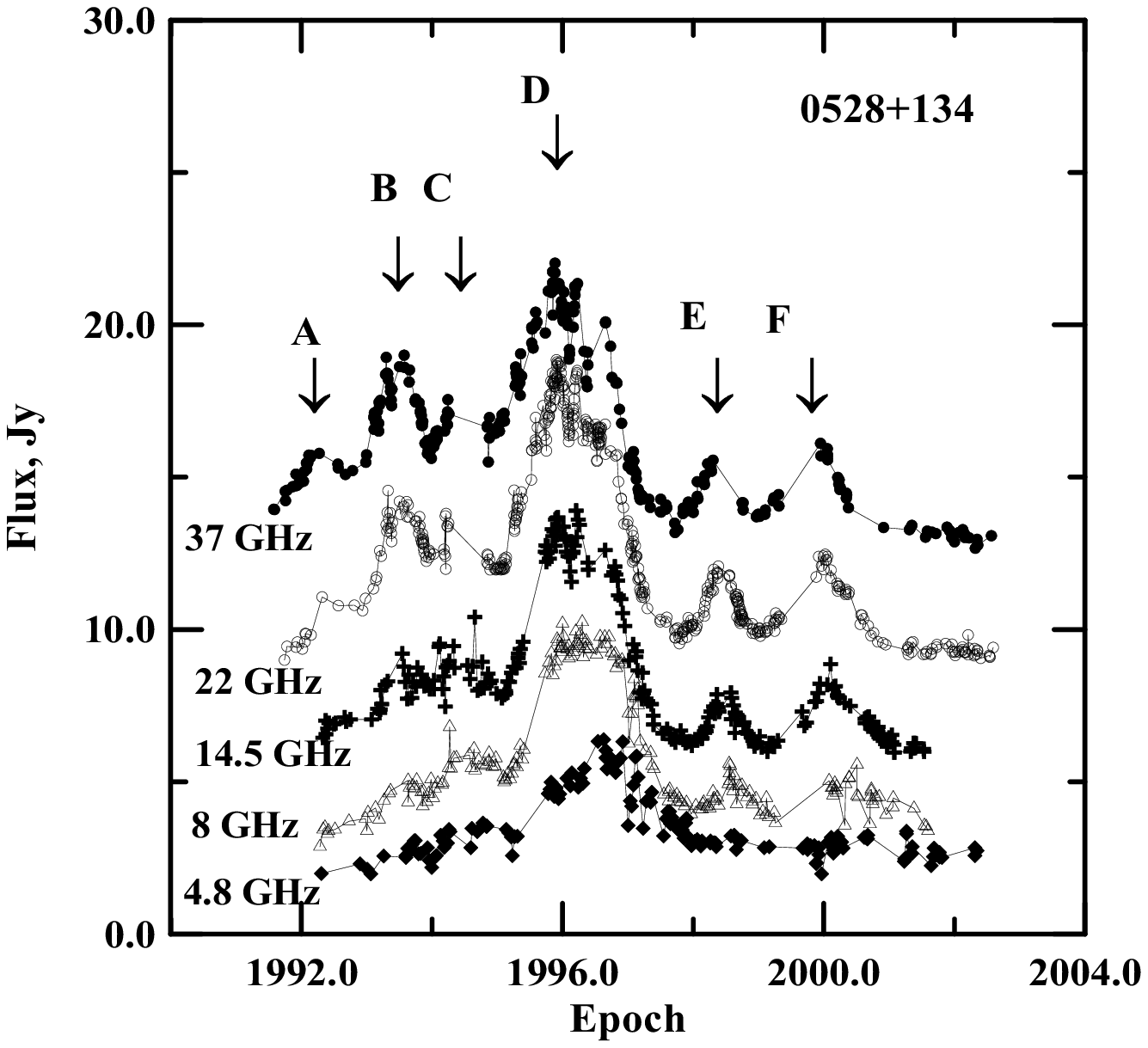}{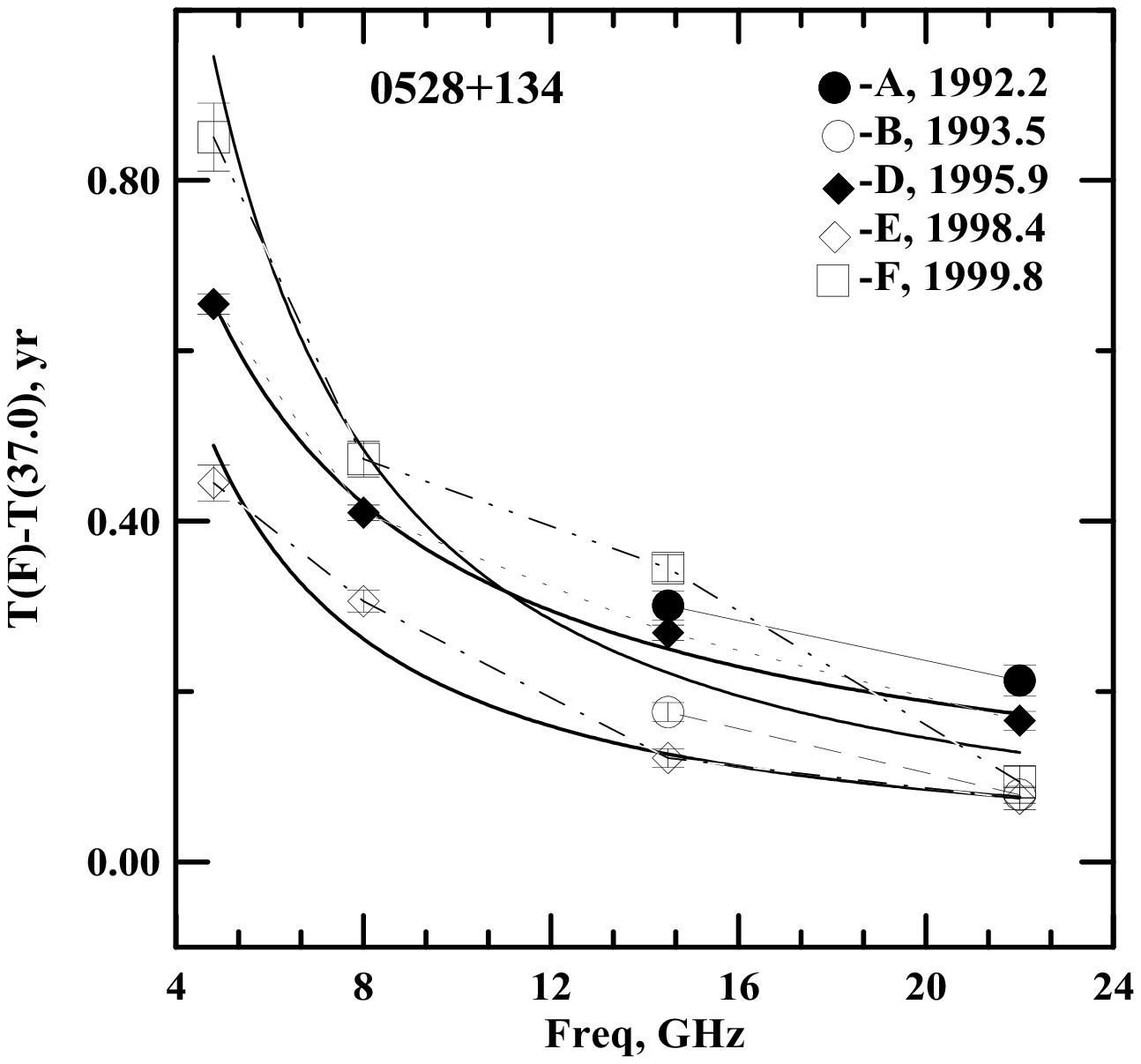}
\caption{({\it Left}) Fine structure of the 1992--1999 core outburst of
0528+134. The arrows with letters mark the positions of individual spikes.
({\it Right}) Time-lags as functions of frequency for the spikes.}
\end{figure}

The light curves of 0458--020 for 1985--2002 display three outbursts
with their maxima near 1989, 1993 and 1995. Only the first of these shows
frequency-dependent time lags and can accordingly be classified as a core
outburst.  The time lag at 4.8~GHz is $(0.93\pm0.03)$~yr, with the index
$\alpha=1.5$. The last two outbursts show no time lags, and can be
considered jet outbursts.

In 1730--130, an extremely bright and narrow
spike (half-width $\le4$~months) was observed at 230~GHz in 1995
(Bower et al. 1997). Weak signs of the spike can also be seen
at 37~GHz, but fade at lower frequencies.
The time delay at 4.8~GHz for the core outburst of 1995--1997
is $(0.93\pm0.02)$~yr, with $\alpha=2.2$.

The outbursts in 0528+134 and 2230+114 display fine structure and
can be resolved into narrow ($<1$ year) spikes.  The time lags and
indices of the exponential functions vary from spike to spike (Fig.~1).
Savolainen et al. (2002) suggested that outburst fine structure
can be induced by shocks that grow and decay in the innermost few
tenths of a milliarcsecond. The bright outbursts in 2230+114 seem to appear
quasi-periodically at intervals of $(8.0\pm0.3)$ yr. The last two of three
observed outbursts are shown in Fig.~2.
Although the amplitudes
of individual spikes vary from one quasi-period to another, their relative
positions are preserved, within the uncertainty introduced by
variations in the time lags.
The first maximum of the next outburst in 2230+114 is
expected near $2005.5\pm0.3$.
\begin{figure}
\plottwo{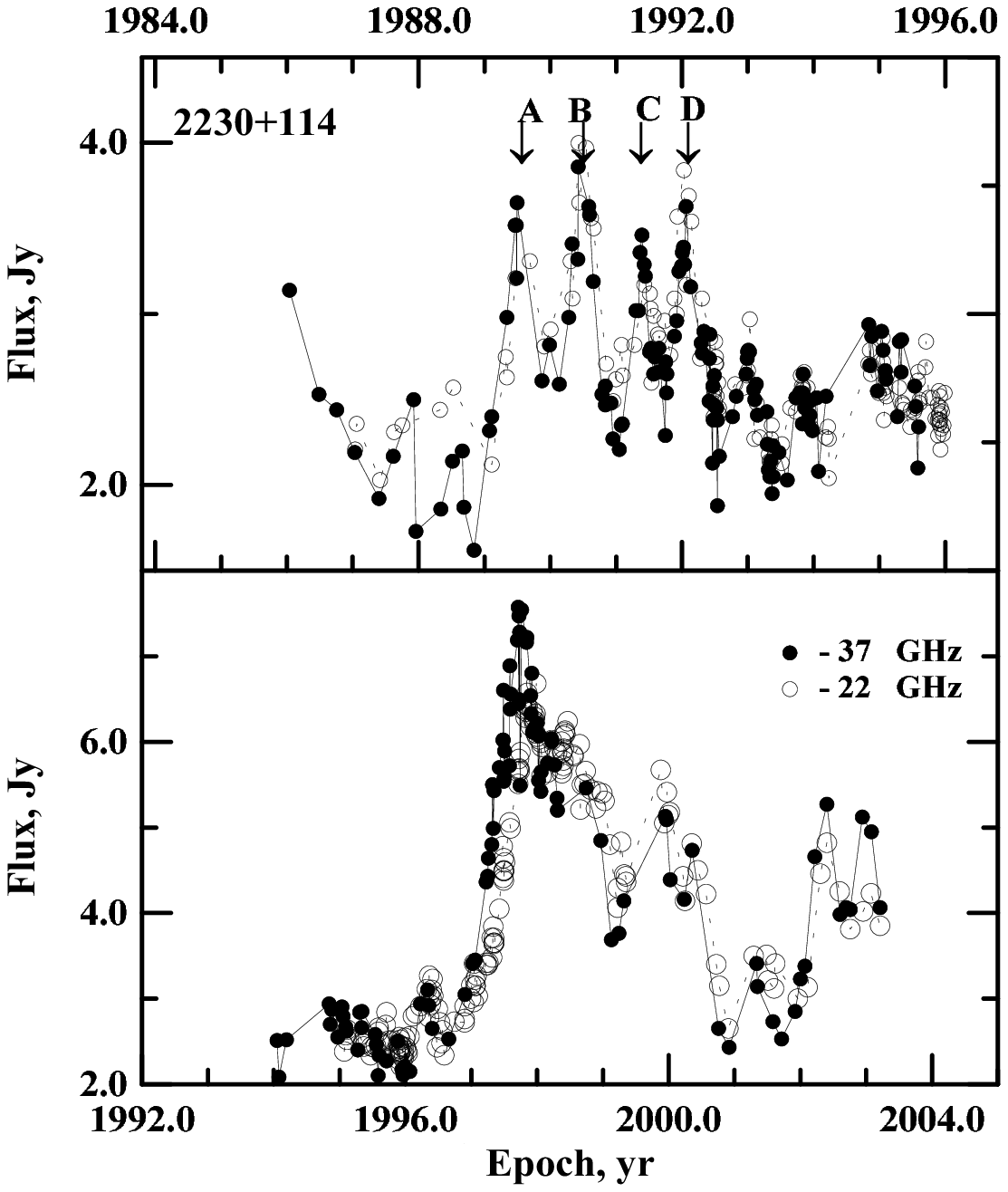}{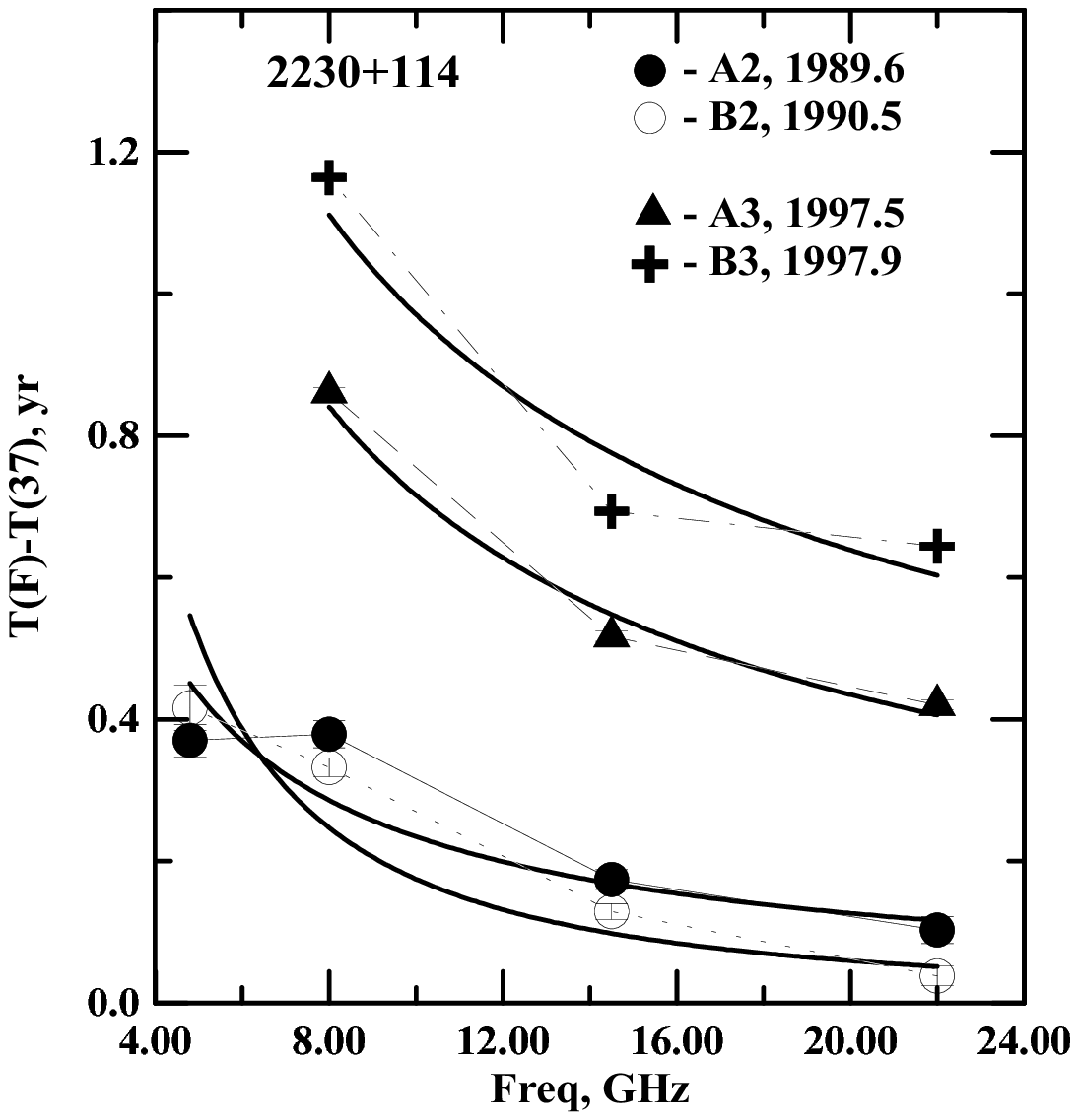}
\caption{({\it Left}) The last two outbursts in 2230+114. The onset of the
second outburst is aligned below the onset of the first, and the arrows
with letters mark the four individual spikes making up each outburst.
({\it Right}) Time-lags as functions of frequency for the spikes.}
\end{figure}
The time lags for the two periods of activity in 2230+114 shown in Fig.~2
are correlated with the corresponding outburst amplitudes: spikes A and B
of the brightest outburst, with its maximum near 1997.5, show greater
time lags than spikes A and B of the outburst with its
maximum near 1989.6.

A powerpoint presentation of this material is available at the web site
http://www.aoc.nrao.edu/events/VLBA10th/posters.html.

\end{document}